# SINGLE LINE APOLLONIAN GASKETS:
# IS THE LIMIT A SPACE FILLING FRACTAL CURVE?

L.M.G. FEIJS

University of Technology Eindhoven and LAURENTIUS LAB. Sittard (The Netherlands)
l.m.g.feijs@tue.nl

ABSTRACT: In this manuscript we study single-line approximations and fractals based on the Apollonian gasket. The well-known Apollonian gasket is the limit case of configurations of kissing circles. Rather than plotting the circles as discs on a differently colored background (the traditional representation), we draw all circles as one line without lifting the pen and without crossing itself. Moreover, the configurations are nested. In this manuscript we explore whether the limit of the line drawings gives rise to a space filling fractal curve.

## 1. Introduction

In their Bridges 2022 paper, Feijs and Toeters [3] develop single-line approximations and fractals based on the Apollonian gasket and apply these in an artistic way for fashion. In [3] the strategies and algorithmic procedures of generating the single-line drawings are described. Visual inspection of the results suggests that the single-line drawings may have a limit, which could be a space-filling curve. It is the purpose of the present manuscript to explore whether this limit does exist and if so whether it is a space-filling curve [6].

The structure of this manuscript is as follows. In Sections 2 we define the single-line approximations of the traditional Apollonian gasket. Moreover, we present refined versions of the gasket, obtained by filling the open areas of the gasket by means of nesting, as proposed by De Comité [2]. The strategies and algorithmic procedures of generating the single-line drawings described in [3] are summarized. Then Section 3 addresses the question whether the successive lines define a curve. Following Sagan [5], a curve is the image of a continuous function $f: [0,1] \to \mathbb{R} \times \mathbb{R}$ using the Euclidean metric. Next, Section 4 explores whether the curve is fractal and space-filling.

## 2. Single-line approximations of the Apollonian Gasket

In this section we summarize the algorithmic procedure of generating the single-line approximations of the traditional Apollonian gasket described in [3]. An example of such a single-line drawing is Figure 1 (left). We call this type I (traditional), to be refined later into a type II (nested). The single line is seen to go round the circles of a typical Apollonian configuration. The numbers refer to the order in which the circles are generated as the Apollonian configuration is produced in a stepwise manner.

We define an *initial configuration* as an ordered set of three distinct circles $c_0$, $c_1$ and $c_2$ such that:
1. $c_1$ and $c_2$ are inside $c_0$,
2. $c_0$ is tangent to $c_1$, $c_0$ is tangent to $c_2$, and $c_1$ is tangent to $c_2$.

We say that the circles $c_0$, $c_1$ and $c_2$ are *mutually tangent*.

---

Next, an *Apollonian configuration* is any ordered set of circles $\{c_0, c_1, c_2, \dots, c_{n-1}\}$ such that:

1. $\{c_0, c_1, c_2\}$ is an initial configuration,
2. Each $c_i$ $(2 < i \leq n-1)$ is obtained from three circles $c_p, c_q, c_r$ $(p \neq q, p \neq r, q \neq r)$ such that $0 \leq p, q, r < i$ and $c_i$ is tangent to all three circles $c_p, c_q, c_r$.

For step 2 it is practical to use the well-known Descartes theorem for finding the radius of $c_i$ from the radii of $c_p, c_q, c_r$. We say that the circles $c_p, c_q, c_r$ are the *parents* of $c_i$. For example, circles 1, 2, and 3 are parents of circle 5 in Fig. 1. Similarly circles 0 (which is not shown in Fig. 1), 3 and 7 are parents of circle 16. In summary, an Apollonian configuration is a set of kissing circles obtained by starting with one outer circle and two arbitrary mutually touching circles inside. Each triple of mutually tangent circles gives rise to two new circles. This stepwise construction process can be repeated, gradually adding more and more circles in the gaps between the already constructed circles.

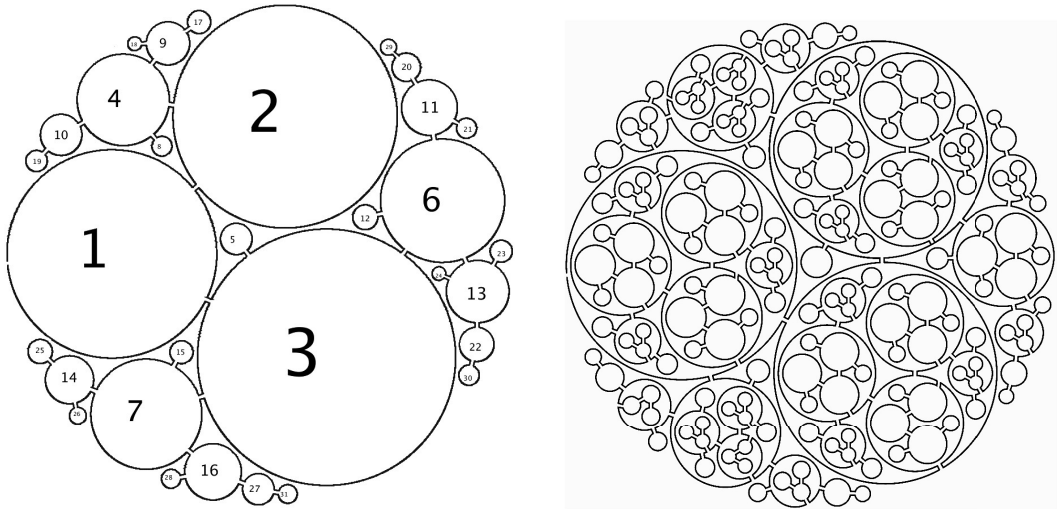

**Figure 1:** *Example of a single-line approximation of the traditional Apollonian gasket (type I, left and type II, nested, right).*

For the construction of an Apollonian configuration we use a radius-based *stop-criterion*, which means that there is a given number $R_{min}$ such that all possible circles $c_i$ whose radius $r(c_i) \geq R_{min}$ are included, and no circles having radius less than $R_{min}$. The effect is a certain uniformity in visual appearance. In practice, the generation algorithm repeats selecting a triple $c_p, c_q, c_r$ of already generated circles, then calculates $r(c_i)$ for the circle (or each of the two circles) tangent to the triple (Descartes' theorem) and then uses an if-clause to only add such new $c_i$ if its radius is large enough. The algorithm stops if no more new $c_i$ can be found.

If we allow the process of refining an Apollonian configuration to go on forever, and then it gives rise to a fractal known as the Apollonian gasket. We generate a single-line approximation of the Apollonian gasket by first generating an Apollonian configuration for given $R_{min}$ (so the configuration approximates the gasket) and then apply a tracing algorithm to the Apollonian configuration. So let us assume we have an Apollonian configuration which includes all possible circles of radius $\geq R_{min}$.

For tracing, different strategies are possible. Many well-known fractals have a tree-shaped hierarchy (Pythagorean tree, Koch curve, Sierpiński triangle), but the Apollonian hierarchy is more complex, as each circle has three parents, not one. In this manuscript we shall restrict ourselves to the so-called hierarchical tracing strategy. In [3], other strategies are discussed, such as the eager (greedy) strategy.

Next, we describe the tracing algorithm. Assume a given Apollonian configuration based on a number $R_{min}$ which is less than the radii of $c_0$, $c_1$ and $c_2$ of the corresponding initial configuration. From the given Apollonian configuration, we construct a continuous line as follows:

1. First, construct for each circle the list of circles touching it, sorted by angular position, going counter-clockwise. We call the circles in this list the *satellites*.
2. Next, choose an entry point for circle $c_1$, say at $-179°$ (almost nine o'clock).
3. Then going round along circle $c_1$ trace its contour, going counter-clockwise, and enter an eligible satellite when passing by at the point of tangency, and as a subroutine trace that satellite too before continuing to trace the rest of circle 1 (and more satellites).
4. Repeat steps 1, 2 and 3 recursively when tracing satellites.

The outermost circle is not traced. We still ought to describe which satellites to trace immediately, and which to skip (in step 3 of the above continuous line construction). Note that a circle appears in at least two satellite lists, some even more. The strategy is hierarchical: break out to *direct children* only (don't skip generations). Thus we need to define the notion of  eligible of tracing step 3.

First, we need to define a concept of generation. For a given Apollonian configuration, for each circle $c_i$ its *generation*, denoted as $\text{gen}(c_i)$. This is defined by:

1. $\text{gen}(c_0) = \text{gen}(c_1) = \text{gen}(c_2) = 0$,
2. For $i > 2$ let $\text{gen}(c_i) = 1 + \max_p \text{gen}(c_p)$ where the $c_p$ are the parent circles of $c_i$.

For example, in Fig.1, circles 0, 1 and 2 have generation 0.

When tracing circle $c$, a satellite circle $c_s$ is said to be *eligible* if:

1. the satellite circle $c_s$ has not been entered before,
2. the generation of $c$ is at most the generation $c_s$,
3. the generations of $c_s$ and  $c$ differ by at most one.

After the continuous line has been found, we choose a number $\delta \leq \frac{1}{4}R_{min}$ and then two more transformations are applied to the continuous line:

1. shrink the circles by a small distance (reducing the radius by $\delta$) ,
2. give a width ($2\delta$) to the entry, so it appears as a bridge connecting circle and satellite.

We call the continuous lines thus obtained an *Apollonian line drawing* or *Apollonian path of type I*. Note that the Apollonian paths do not self-intersect. Next we develop a more sophisticated version of these paths, which we call *type II*, and which is based on recursive circle packing applied to the interiors of the existing circles as in [2] and [3]. The procedure to generate paths of type II will be outlined in the remainder of this section. We apply recursive circle packing to the interiors of the existing circles as in [2], next apply line tracing again. For the nesting we restrict ourselves to circle configurations in which the three initial circles have the same size.

A *nested Apollonian configuration* is like an Apollonian configuration, but now some of the circles $c_i$ carry with them another Apollonian configuration with  $c_i$ being its outer circle, and so on, recursively.

We embed the recursive tracing of the interiors into the overall tracing of the circles and the satellites. The tracing algorithm for nested Apollonian configurations is like the algorithm given in the previous section, with two adaptations:

1. When a circle is large enough to accommodate an interior, choose a suitable angular position at the circle's circumference for turning inward. This position should not be close to an eligible satellite nor close to the entry (where drawing the circumference began). Solution: we look for a suitable inward-turning point by first trying the best options, $-60°$ and $180°$ ($0°$ means 3 o'clock). Then the nearest of the main interior circles is entered. If these options are unavailable, divert to positions near $-60°$, for example $-63°$ (allowing the bridge to become slanted).
2. Tracing circles in the interior goes clockwise if the circle itself goes counter-clockwise (default), otherwise the turning-inward bridge would become an ×.

The resulting nested Apollonian paths of type II are shown in Fig. 2 (from [3]). The six paths shown in Fig. 2 are obtained by taking smaller and smaller values for $R_{min}$.

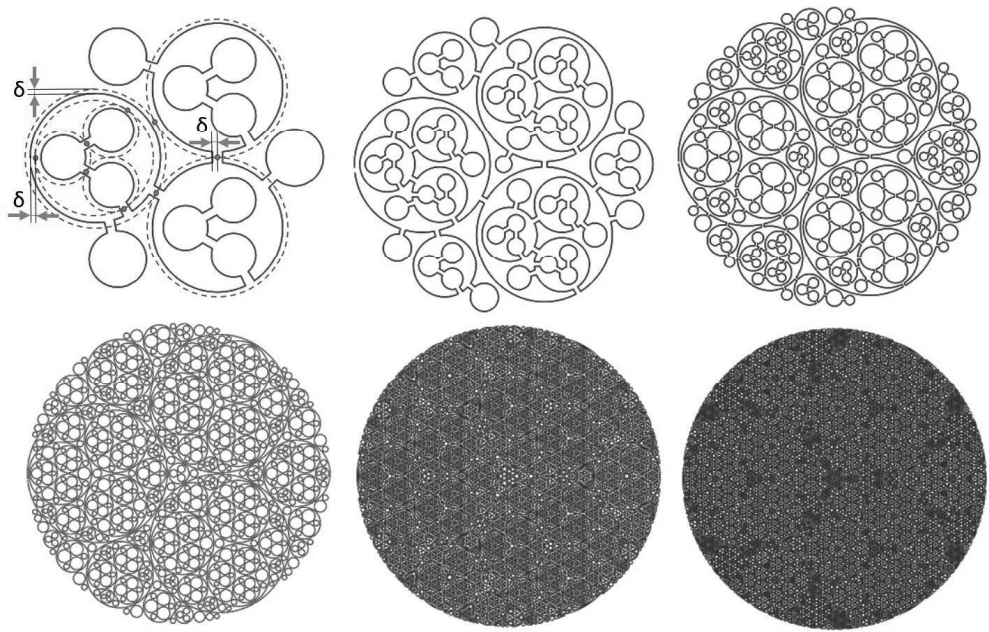

**Figure 2**: *Nested Apollonian paths: successive versions with smaller minimal radius.*

A key question is whether such successive paths (of type II) converge to a curve which is space-filling, as suggested by the visual appearance of the successive paths in Fig. 2. There are several questions: Can we define a limit construction and if so, is it a curve? Does it have fractal nature? Does it fill the space inside the outer circle? These are the main questions of the present manuscript, addressed in the next sections.

## 3. Is it a Curve?

Sagan [6] defines a *curve* as the image of a continuous function $f:[0,1] \to \mathbb{R} \times \mathbb{R}$ using the Euclidean metric. We have type II paths for $n = 1, 2, ...$, the outcomes of the tracing algorithm applied to the $n$-th approximation of the gasket. We let $R_{min}(n) =$ "constant" $\times r^n$ where $0 < r < 1$ (we used $r = \frac{1}{2}$ for Fig. 2, $r = \frac{1}{2}\sqrt{2}$ for Fig. 5). Shrink distance and half bridge width depend on $n$ by $\delta = \frac{1}{4} R_{min}$.

Since we build with line segments and arcs, each $n$ produces a continuous path with length, $l(n)$ say. As a first idea, define $f^n:[0,1] \to \mathbb{R} \times \mathbb{R}$, mapping $t$ to the point on the path at distance $t \times l(n)$ from the beginning. If $\lim_{n\to\infty} f^n(t)$ exists, let $f(t) = \lim_{n\to\infty} f^n(t)$. But does the limit exist? New satellites change the handling order. Smaller $R_{min}$ enable interiors which consume extra length.

We sketch how to define a curve: during tracing, when the interior of a circle has to be done, the three largest circles are traced, the 2nd being traced in a subroutine from the 1st etc. (Fig. 3, left). Thus four "triangular" areas appear, each of which must be colonized (traced) in subroutines. Whenever a triangular area has to be done, it is by a central circle from which three new triangular areas are colonized (Fig. 3, 2nd). Thus circles arise with 0, 1, 2 or 3 subroutines (next to an inward subroutine).

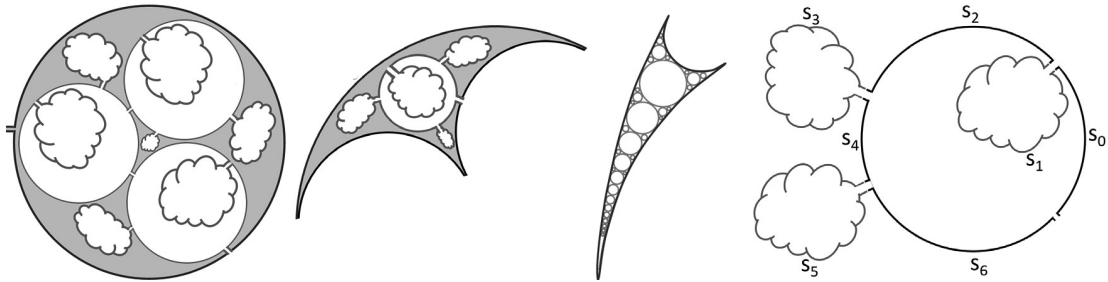

**Figure 3**: *Colonizing areas from circles, and splitting a circle's path into segments*

Sagan (p. 11) describes the construction of the Hilbert curve: "every $t$ in the closed unit interval is uniquely determined by a sequence of nested closed intervals (that are generated by our successive partitioning), the lengths of which shrink to 0. With this sequence corresponds a unique sequence of nested closed squares, the diagonals of which shrink into a point." Here we do the same, recursively partitioning the unit interval into 3, 5, 7, or 9 sub-intervals. The number of intervals follows the subroutine branching, which splits each circle's path into what we call *segments*. In Fig. 3 (right) we see a circle's path being split in seven segments, which are either simple arc-segments ($s_0$, $s_2$, $s_4$, $s_6$), or circle-segments, which can still expand ($s_1$, $s_3$, $s_5$). Instead of Hilbert's squares, we have *areas* of several types: disks and triangles for circles whose segment count is not final, and arc-segments for circles with final segment count. A noteworthy triangle area is the spikey horn, shown third in Fig. 3.

Consider the five segments of the first circle's path in Fig. 4 (left). If we want to map a value from $[0,1)$, say $t = 0.23$ to one of these segments, we divide the interval $[0,1)$ into five same-length intervals $[0,\frac{1}{5})$, $[\frac{1}{5},\frac{2}{5}),..,[\frac{4}{5},\frac{5}{5})$. Then 0.23 is mapped to $s_1$, since $0.23 \in [\frac{1}{5},\frac{2}{5})$. A circle's segment count can still increase (as in Fig 8), but that will take no longer than $\lceil 1.87/\ln r \rceil$ steps[2] and is bounded by 9. The configurations obtained by further refinement are stable: the sudden appearance of extra path-length in a subroutine will not disturb points elsewhere.

The $n$-th path $f^n$ is a continuous function of $t$. For each $t$, $\lim_{n\to\infty} f^n(t)$ exists, roughly speaking because the size of the areas (disks, triangles, arc-segments) around $f^n(t)$ can be made arbitrarily small. There is a complication: when $n \to \infty$, the areas move a bit because of the *shrink distances* and the *bridges*. Therefore, the environments (of radius $< \epsilon$) required for the limit construction are not these areas, but open discs around them, after widening by $\delta \times$ "recursion depth". The environments around $f(t)$ shrink toward the limit point by two effects. The first is the recursive partitioning of disks, triangles and arcs, which gives arbitrarily small areas. The second is the vanishing of (stacked) shrink distances and bridges. Note that $\delta \times$ "recursion depth" $\leq \delta \times n = \frac{1}{4}R_{min} \times n \propto r^n \times n \to 0$ when $n \to \infty$. Convergence is uniform[3], so $f$ is

---

[2] The satellites are of the same order of magnitude as the circle itself. The *worst case* satellite is amidst three equal-sized large circles, in which case their ratio is $3 + 2\sqrt{3} = 6.464$. When $r = 2$, for example, any eligible satellite will thus pop up within three steps. We used that $\ln 6.464 = 1.87$.

[3] The sequence $f^n$, $n = 1,2,3,...$ is said to be *uniformly* convergent if, for each $\varepsilon > 0$ an integer $N$ can be found such that $|f^n(t) - f(t)| < \varepsilon$ for $n \geq N$ and all $t$ in the given interval. So if we want all (widened) enclosing areas to fit in discs of radius $< \varepsilon$, then we choose an $N$ such that even the most difficult triangular area is (enclosed in a disc of radius) less than $\varepsilon$. The spikey horn of Fig. 3 is the most difficult, but for any $\varepsilon$ we can develop enough circles deep down along the necklace so that only a tip of

continuous (the uniform limit of continuous functions remains continuous). Although the definitions are sketchy, we conjecture that the paths define a curve.

Sagan [6] (p.14) presents a quaternary representation of $t$ for the Hilbert curve. We could do a similar coding, using mixed-base (ternary, 5-ary, 7-ary, etc.). The unfolding of paths yields the base sequence.

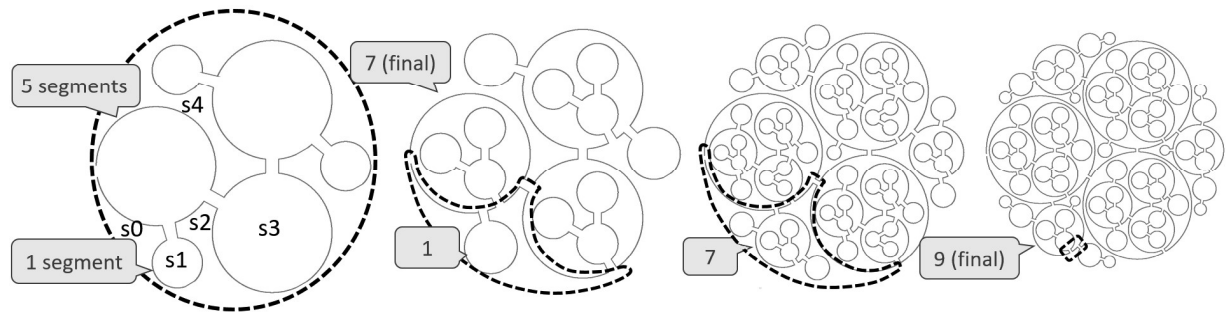

**Figure 4**: *Areas around $f^n(t)$ for $t = 0.23$ in paths various refinement levels.*

## 4. Is it a Space-filling Fractal Curve?

Are the curves of type I and type II defined in Section 2 a fractal? We used our program to generate paths for various values of $R_{min}$ from 128 down to 2 pixels in 12 steps and found total lengths ranging from 3334 to 285736 (pixels), which appear as an almost straight line when plotted in a log-log manner (the upper line in Fig. 5, right).

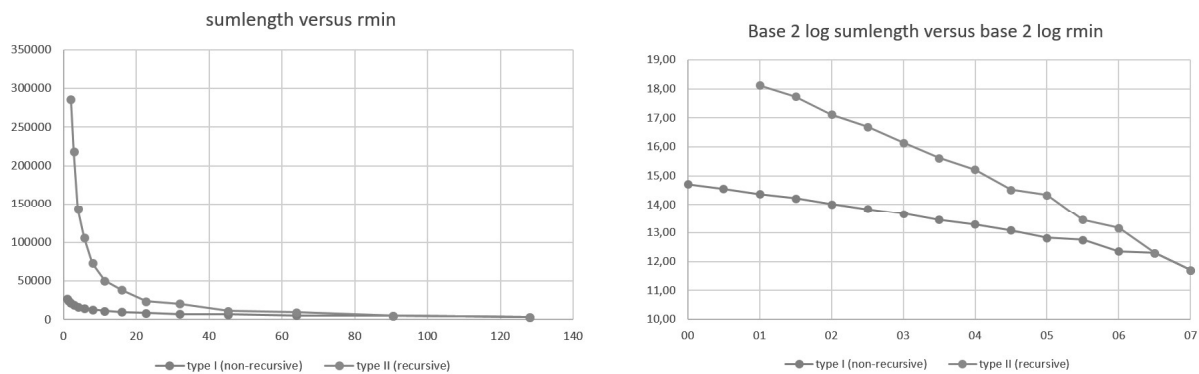

**Figure 5**: *Plot of summed path length versus $R_{min}$ and base 2 log-log plot of the same.*

Therefore the curves of Section 2 have fractal nature (which is no surprise, the fractal nature is inherited from the Apollonian gasket and the recursive nesting). For type II, the slope of this log-log plot for $2 \leq R_{min} \leq 64$ is $-1.011$ (SPSS Linear regression, Standard Error 0.020) which suggests a fractal dimension close to 2. Theoretically the fractal dimension should be 2, because eventually, the successive paths will come arbitrarily close to any point inside the outermost circle. This also suggests an answer to the question whether our fractal curve is space-filling: we conjecture it is.

---

the horn shorter than $2\varepsilon$ remains. If we do this for the big horns of the outermost disk, then all other triangles, including those in nested recursions, are smaller. Therefore one value of $N$ does the job for all $t$.

## 5. Related work

In The Fractal Geometry of Nature [1], Benoît Mandelbrot writes that the Sierpiński gasket can be knitted with a single loop of thread and that the same is true of the Apollonian net. Mandelbrot presents neither full details of the loop, nor needlework. In Indra's Pearls [5], Mumford, Series and Wright generate fractal loops by iterating Möbius transforms. Inglis and Kaplan [4] and De Comité [2] place small circles inside the large ones. The fractals of [2] are called "avatars of the Apollonian gasket". Feijs and Toeters [3] develop single-line approximations and fractals based on the Apollonian gasket, coded the tracing in Processing.

*Acknowledgements*: The author likes to thank Marina Toeters and all colleagues of the Fashion Tech Farm in Eindhoven for the support and cooperation on the topic of this manuscript.